\documentclass[aps,prl,twocolumn,groupedaddress]{revtex4-1}
\usepackage[english]{babel}
\usepackage[utf8]{inputenc}
\usepackage[colorinlistoftodos, color=green!40, prependcaption, textwidth=0.7in]{todonotes}

\usepackage[pdftex, pdftitle={Article}, pdfauthor={Author}]{hyperref} 

\usepackage{hyperref}       
\usepackage{url}            
\usepackage{amsfonts}       
\usepackage{nicefrac}       
\usepackage{microtype}      
\usepackage{physics}
\usepackage{amsmath}
\usepackage{subfigure}

\usepackage{xcolor}

\graphicspath{ {./img/} }

\begin{document}
\title{Unitary long-time evolution with quantum renormalization groups and\\ artificial neural networks}

\begin{abstract}


In this work we combine quantum renormalization group approaches with deep artificial neural networks for the description of the real-time evolution in strongly disordered quantum matter. We find that this allows us to accurately compute the long-time coherent dynamics of large, many-body localized systems in non-perturbative regimes including the effects of many-body resonances. Concretely, we use this approach to describe the spatiotemporal buildup of many-body localized spin glass order in random Ising chains. We observe a fundamental difference to a non-interacting Anderson insulating Ising chain, where the order only develops over a finite spatial range. We further apply the approach to strongly disordered two-dimensional Ising models highlighting that our method can be used also for the description of the real-time dynamics of nonergodic quantum matter in a general context.

\end{abstract}

\author{Heiko Burau}
\email{burau@pks.mpg.de}
\author{Markus Heyl}
\affiliation{Max-Planck-Institut f\"ur Physik komplexer Systeme, N\"othnitzer Straße 38, 01187 Dresden, Germany}

\maketitle

\textit{Introduction.} The understanding of emergent behavior in quantum many-body systems is largely based on the discovery of effective descriptions of analytically unsolvable models \cite{anderson1972more}. 
An essential toolkit to find the former constitute renormalization group (RG) methods.
They are traditionally applied on systems in thermal equilibrium, thereby explaining many collective phenomena including structured phases, phase transitions, critical scaling and universality. 
In the past decade, real-space RGs have been developed that aim to explain analogues of these well-known phenomena also in systems where a thermodynamic treatment breaks down due to strong quenched disorder \cite{anderson1958absence, basko2006metal, gornyi2005interacting, pal2010many, oganesyan2007localization, nandkishore2015many, smith2016many, protopopov2020non, rademaker2016explicit}. 

Whereas real-space RGs successfully operate in the stationary setting at the level of individual eigenstates~\cite{vosk2013many, pekker2014hilbert, altman2015universal, morningstar2019renormalization, dumitrescu2019kosterlitz, goremykina2019analytically},  reaching a quantitative description of the \textit{dynamical} properties of quantum many-body systems appears even more challenging.
So far, coherent dynamics of quantum matter far from equilibrium has been  mostly simulated using tensor networks methods~\cite{vznidarivc2008many, andraschko2014purification, vidal2003efficient} or exact diagonalization~\cite{luitz2016extended, kollath2007quench} with recent developments targeting dynamical descriptions in terms of machine learning methods by utilizing Restricted Boltzmann Machines (RBM) \cite{carleo2017solving} or, more general, Artificial Neural Networks (ANN) \cite{carleo2017solving,schmitt2019quantum, schmitt2018quantum}. 
Still, accessing quantitatively the long-time dynamics for large quantum many-body systems, especially in spatial dimensions beyond one, represents a major challenge~\cite{guardado2018probing, de2019efficiently, hackl2008real}.
\begin{figure}[t]
	
	\vspace{-4mm}
	
	\includegraphics[width=1\columnwidth]{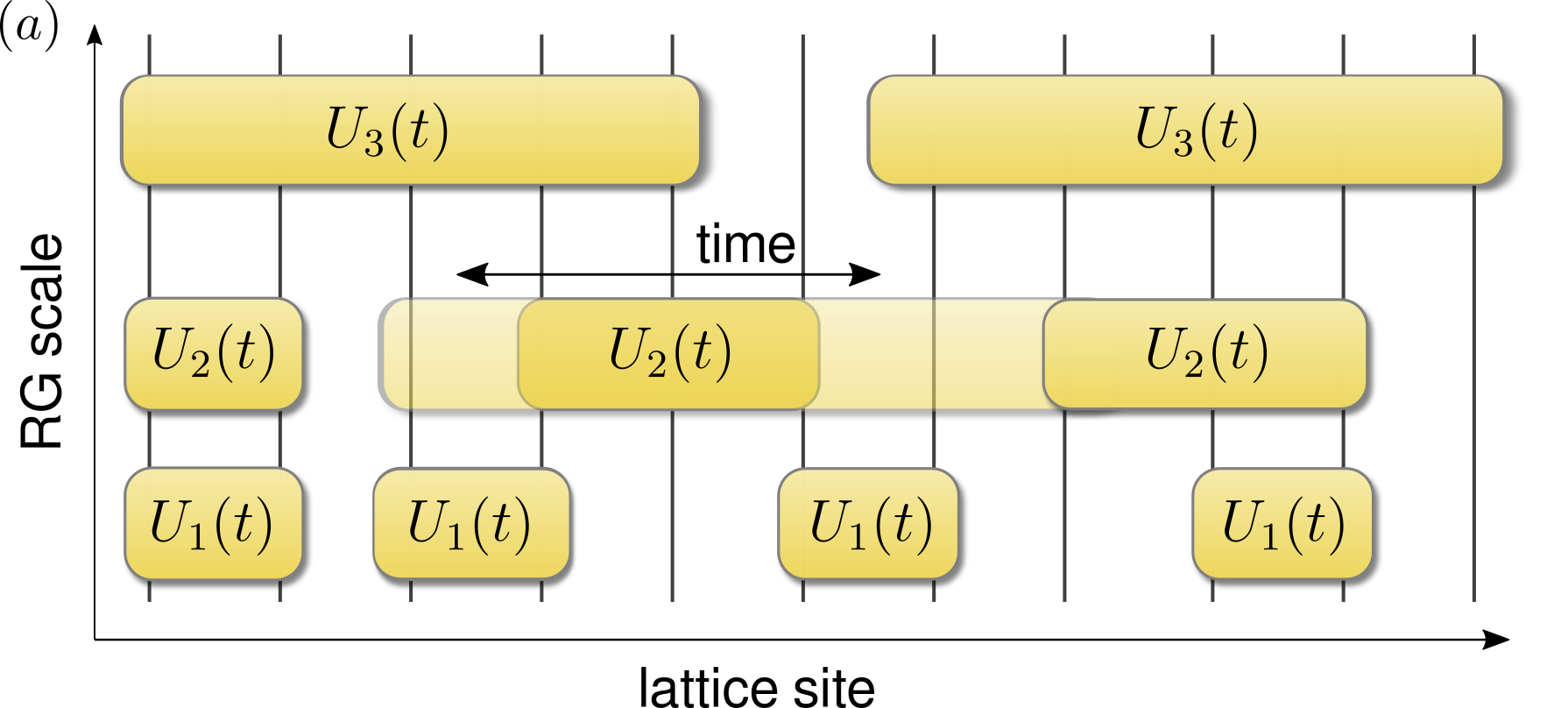}

	\vspace{3mm}
	
	\includegraphics[width=1\columnwidth]{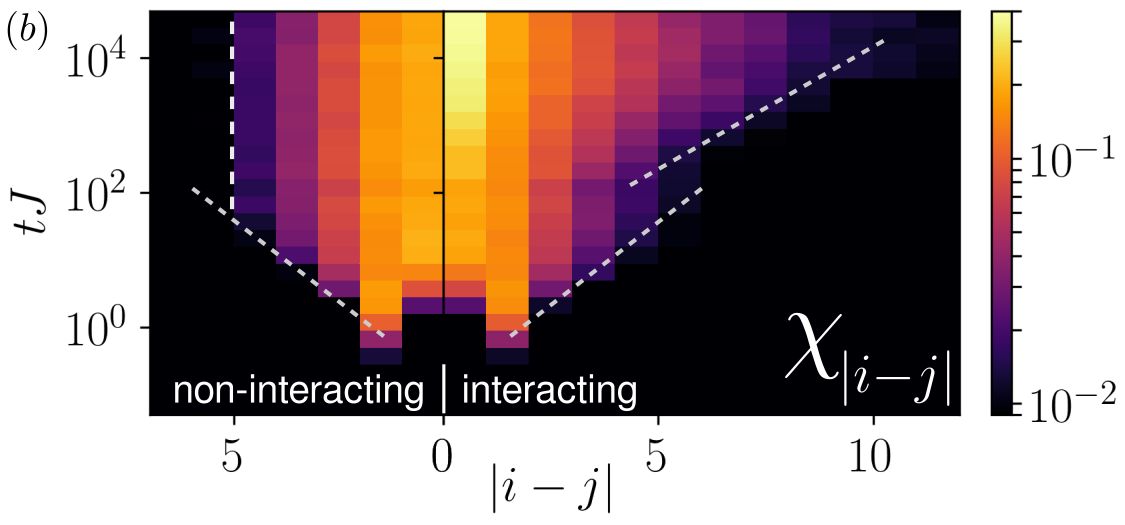}
	
	\caption{(a) Illustration of a random quantum circuit built up from local unitary RG-transformations. In the course of the RG, long-distance and higher-order couplings emerge. Adding time-dependence leads to a further broadening with increasing time. (b) Spatiotemporal build-up of MBL spin-glass order in a random quantum Ising chain with $64$ lattice sites after quenching a paramagnetic initial condition into the symmetry-broken phase at $J = 5 h, J^{(x)} = h / 5$. Dashed lines indicate emergence of light-cones except for the non-interacting case, $J^{(x)} = 0$, where the order stops developing at a finite distance. The numerical data is obtained from an average over 25 disorder realizations.}
	
	\label{fig:spin_glass_order}
	\label{fig:quantum_circuit}
	
\end{figure}

In this work, we show how ANNs can be utilized in a different way for numerically exactly time-integrating effective descriptions of generically interacting systems, generated by RG methods.
As a concrete example, we explore the temporal build-up of MBL spin-glass order out of a simple polarized state for a large, disordered spin chain, see Fig. \ref{fig:spin_glass_order}b, among other long-time dynamics in 1D- and 2D-lattices.
We begin by formulating a prototypical strong-disorder RG (SDRG) for spin-1/2 systems of arbitrary spatial dimension
and map its transformations into the time-domain.
As a result, we obtain a quantum circuit, see Fig. \ref{fig:quantum_circuit}a, as an effective description of the time-evolution operator. 
Hereafter, we show that this circuit can be encoded efficiently into deep ANNs associated with typical initial conditions for quantum real-time dynamics.  
This allows us to quantitatively represent time-evolved many-body quantum states not only at short but also long times.
We note that our method avoids a discretization of time but relies on a renormalized Hamiltonian that is assumed to effectively describe the relevant physics up to some finite but nevertheless long time scale.

The scheme we introduce in the following can be applied for any generic, but strongly disordered, spin-1/2 system.
Concretely, we will apply it to a paradigmatic interacting disordered quantum Ising model \cite{fisher1995critical} of the form
\begin{equation}\label{eq:model}
	H = \sum_{\langle ij \rangle} J_{ij} \sigma^z_i \sigma^z_j + J^{(x)}_{ij} \sigma^x_i \sigma^x_j +\sum_i h_i \sigma^x_j ,
\end{equation}
with next-neighbor couplings $J_{ij} \in [-J, J]$ and local magnetic fields $h_i \in [-h, h]$ drawn randomly from uniform distributions. 
We use periodic boundary conditions.
For the 1D case we also add random transverse couplings $ J^{(x)}_{ij} \in [-J^{(x)}, J^{(x)}]$ to obtain a generic and interacting model.

\textit{Solving the time evolution.}
Before describing the utilized renormalization procedure and the training of the ANN in detail, let us start by outlining the general scheme for solving quantum real-time evolution utilizing strong-disorder RGs.
Such an RG generates a sequence of local unitary transformations $U_k$ in order to iteratively obtain a simplified effective description of the considered quantum many-body system.
In the time domain, we will show that this leads to the following representation of time-evolved quantum many-body states: 
\begin{equation}\label{eq:psi_t}
	\ket{ \psi(t) }_{QC} = e^{-i H_0^{(n)} t}\ U_1(t) \cdots U_n(t) \ket{ \psi_0 } \, ,
\end{equation}
where a time-dependence is added to the RG-trans\-formations $U_k$ through a generalized interaction picture, see the derivation below.
%
The above equation maps quantum dynamics onto a quantum circuit generated by the local unitaries $U_k(t)$.
As we assume that the effective description in terms of the final Hamiltonian $H_0^{(n)}$ after the end of the RG procedure can be solved exactly, the complexity of the quantum circuit emerges solely unitaries $U_k(t)$.
We find that such quantum circuits can become a non-perturbative object, as the spatial support of the $U_k(t)$ typically grows over time developing long-distance and higher-order couplings with large overlaps, see Fig \ref{fig:quantum_circuit}a. 
A central contribution of this work is to outline a numerically exact scheme to encode $\ket{ \psi(t) }_{QC}$ and therefore the RG transformation itself into an ANN using machine-learning techniques. 
The numerical learning effort in obtaining $\ket{ \psi(t) }_{QC}$, as well as its memory requirement, scales at most quadratically with system size while being independent on the targeted time $t$ or the spatial dimension.

\textit{Dynamical strong-disorder Renormalization Group.}
In principle, quantum circuits such as in Eq.~(\ref{eq:psi_t}) can be generated using a variety of standard SDRGs.
In the following we introduce a variant of a SDRG, which as we find improves the quantitative accuracy of the resulting scheme.

As other SDRGs, the dynamical variant we introduce is based on a local separation of energy scales.
Consequently, at the beginning of each iteration $k$ we pick the strongest coupling, also called "fast mode", whose corresponding term in the Hamiltonian we call $H_0$.
For the first iteration this could be either a spin interaction $J_{ij}$, $J_{ij}^{(x)}$, or transverse field $h_{i}$, see the Hamiltonian in Eq.~(\ref{eq:model}).
Those terms in the Hamiltonian which are not commuting with $H_0$ we denote by $V$.
These can be eliminated perturbatively using a Schrieffer-Wolff transformation (SWT) \cite{bravyi2011schrieffer} by applying a unitary transformation $W_k = e^{S_k}$ on the Hamiltonian with a generator $S_k$ satisfying $[H_0, S_k] = V$ and $S_k^\dagger = -S_k$ \cite{pekker2014hilbert}, at the expense of the renormalization $H_0 \mapsto H_0 + [S_k,V]/2$.
In general, this modifies existing couplings and leads to the generation of new terms in the Hamiltonian.
After the SWT the fast mode is decoupled from the remainder and can then be faithfully removed from the system as a second-order local integral of motion (LIOM) \cite{ros2015integrals, imbrie2017local, rademaker2016explicit}.
After $n$ such iterations, an unperturbed Hamiltonian $H_0^{(n)}$ is obtained, formed by the set of LIOMs.

The newly generated couplings after each iteration are, of course, not known a-priori, especially if the SDRG is designed regardless of details of the model like range of interaction, dimensionality etc. 
We approach this problem by represent at each stage of the RG the Hamiltonian as a sum of \textit{arbitrary} Pauli-strings $\sigma_{l_1}^{\alpha_1} \dots \sigma_{l_M}^{\alpha_M}$ with a real coefficient $\lambda_{l_1,\dots,l_M}$ each.
Certainly, this approach can entail a costly handling of numerous generated higher-order couplings, see below, but it opens the possibility to take into account many-body resonances, which are neglected using earlier SDRGs \cite{vosk2013many, vosk2014dynamical} and related, so-called flow equation approaches \cite{thomson2018time, thomson2020dynamics, hackl2008real}.

In addition the accuracy of the RG can be further increased by splitting the SWT into infinitesimal unitary transformations, closely resembling in spirit the flow equation framework.
This turns out to be particularly helpful in the vicinity of a critical point, $h \approx J$ here for the 1D model, where the SWT is least controlled and in order to capture many-body resonances to an arbitrary degree.
For a detailed presentation of the technical details, see the appendix.
To control the exponential number of couplings $\{V_i\}$ generated during the RG, we first neglect those terms where $|V_i| \ll {t^*}^{-1}$ which are much smaller than the inverse of the targeted time scale $t^\ast$ say, as they do not influence physics up to $t^*$.
Secondly, we perform the continuous renormalization only w.r.t. those $V_i$ whose relative magnitude lies above a fixed threshold, $|V_i| / |H_0| > \epsilon \ll 1$.
Therefore we have a tradeoff, that is controlled by $\epsilon$, between exactness and total number of couplings within the RG-generators $S_k$ and the renormalized Hamiltonian $H_0^{(n)}$.
In our computations, $\epsilon$ typically ranges from $10^{-4}\ldots 10^{-2}$, depending on the closeness to the critical point, $h \approx J$, or the ergodic transition, $J^{(x)} \approx J$.
Later we will present a quantitative analysis of our RG w.r.t. the dynamics of local observables.

\textit{Time-dependent unitaries.}
To derive the time-dependence of $U_k(t)$ we express the time-evolution operator in the renormalized basis, which yields
\begin{equation}
\begin{split}
e^{-i H t} &= e^{S_1^\dagger} \cdots e^{S_n^\dagger}  e^{-i H_0^{(n)} t} e^{S_n} \cdots e^{S_1} \\
&= e^{-i H_0^{(n)} t} e^{S_1^\dagger(t)} \cdots e^{S_n^\dagger(t)} e^{S_n} \cdots e^{S_1} \, .
\end{split}
\end{equation}
We achieve a much more robust learning of the ANN upon successively commuting each factor $e^{S_k}$ to the left until its counterpart $e^{S_k^\dagger(t)}$ is reached. Identifying $U_k(t) = e^{\tilde S_k^\dagger(t)} e^{\tilde S_k}$ gives then the desired form as in \eqref{eq:psi_t}.
Here, $\tilde S_k$ denotes the total application of all rotations from $e^{S_l}, l < k$ on $S_k$, see the appendix for details.

\textit{Training the artificial neural network.}
Utilizing ANNs as a variational ansatz for many-body wavefunctions has seen an active development recently \cite{carleo2017solving,ryczko2019deep} becoming competitive with or partially even superior to other state-of-the-art methods.
%
\cite{schmitt2019quantum, hibat2020recurrent, nagy2019variational}.
%
In contrast to the commonly used time-dependent variational principle (TDVP), we introduce another way of training an ANN.
As such, the scattering operators $U_1(t), \ldots, U_n(t)$ are consecutively used during $n$ iterations to train the network. 
As the $U_k(t)$ are still local operators with a finite support in real space, we perform for each iteration $k$ a supervised learning procedure to find the set of complex network parameters $\mathcal{\tilde W}^{ (k) }$ that minimize the Fubini-Study metrics, given by $L [ \mathcal{\tilde W}^{ (k) } ] = \acos( | \bra{ \psi_{ \mathcal{\tilde W}^{ (k) } } } U_k(t) \ket{ \psi_{ \mathcal{W}^{ (k) } } } |^2  ).$
whereas $\ket{ \psi_{\mathcal{W} } } = \sum_{ \{ \vec s \} } \exp[ H_{\rm ANN}( \mathcal{W}, \vec s ) ] \ket{\vec s}
$ refers to a quantum state defined by the output of an ANN and $\{ \vec s \}$ denotes the set of all spin configurations $\vec s = (s_1, s_2, \ldots), s_i = \pm 1$. Notice that we assume always properly normalized wave functions.
The network $H_{\rm ANN}( \mathcal{W}, \vec s )$ can be considered as a deep extension of a complex-valued RBM with up to three hidden layers, see the appendix for details.
After convergence, the "learned" solution $\mathcal{\tilde W}^{ (k) }$ is passed to the next iteration as $\mathcal{W}^{ (k+1) }$.
To complete the learning procedure, we write $L [ \mathcal{\tilde W}^{ (k) } ]$ at the $k$-th iteration, while omitting the index, as
$L [ \mathcal{\tilde W}^{ (k) } ] = \sum_{ \{\vec s \} } |\psi(\vec s)|^2  \tilde\psi^*(\vec s) [\psi^*(\vec s)]^{-1} U_{\rm loc}(\vec s, t), \ U_{\rm loc}(\vec s, t) = [ \braket{\vec s}{\psi_{\mathcal W} }]^{-1} \bra{\vec s} U(t) \ket{\psi_{\mathcal W}} $ ,
with $\psi(\vec s) = \braket{ \vec s }{ \psi_{\mathcal W} }$ and $U_{\rm loc}(\vec s, t)$ being the equivalent of the local energy known from TDVP methods. 
We access the above sum with a Markov chain Monte-Carlo (MCMC) algorithm which, as we can confirm, is sign-problem free in all our computations. 
In the same way, we calculate the gradient $\partial L / \partial \mathcal{\tilde W}_i$ with the backpropagation algorithm and pass the result to a stochastic gradient-descent optimizer referred to as PADAM \cite{chen2018closing}.

\begin{figure}
\subfigure{
	\includegraphics[width=1\columnwidth]{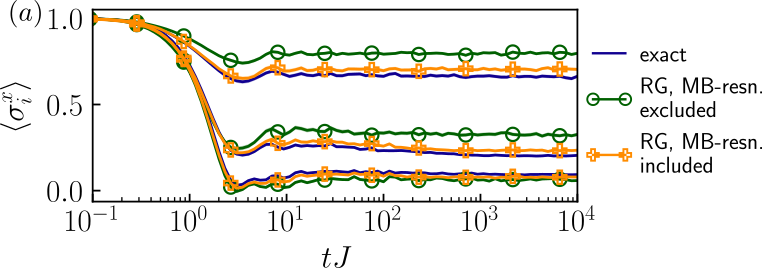}
}
\vspace{-0.3cm}
\subfigure{
	\includegraphics[width=1\columnwidth]{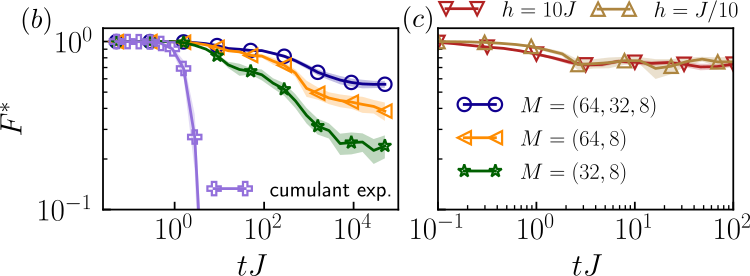}}
	\caption{(a) Comparing the dynamics of the transverse magnetization to exact diagonalization, averaged over lattice sites and 250 disorder realizations, with and without treatment of many-body resonances. Here, $L = 12, h = \{ 4, 1, \nicefrac{1}{4} \} J$ from top to bottom, and  $J^{(x)} = J / 8$. (b) Lower bound $F^\ast$ on the many-body overlap of the trained ANN-state with the state given by an hypothetical, exact application of the quantum circuit. System parameters are the same as in Fig.~\ref{fig:spin_glass_order}b in the interacting case. Different sizes of ANNs are compared, where $M$ refers to the number of hidden units in each layer. Shaded areas indicate uncertainties due to a finite disorder ensemble of 25 realizations. The result from a cumulant expansion of the quantum circuit is shown for comparison. (c) The same overlap on a $12\times 12$ lattice for two different external field strengths.}
	\label{fig:benchmarks}
\end{figure}

\textit{Benchmarking.} In order to quantify the overall accuracy of our approach we first benchmark the RG-component and the machine learning part individually.
For the former task, we calculate Eq.~\eqref{eq:psi_t} for small system sizes exactly using a matrix representation of the quantum circuit.
Fig. \ref{fig:benchmarks} shows a comparison of the local magnetization with the result obtained from exact diagonalization for a system of $L=12$ spins.
The plot reveals that the accuracy of the dynamics depends crucially on the inclusion of many-body resonances, which is tuned by the only free RG-parameter $\epsilon$, see above.
For practical purposes, we set $\epsilon$ indirectly by imposing a maximum total number $\mathfrak n$ of couplings within all RG-generators $S_k$.
Here, $\mathfrak n = 3(10)L$ corresponds to the label of excluded (included) many-body resonances and matches (exceeds) the number of original couplings.
Already for $\mathfrak n=10L$ we observe a very good agreement even for the longest times. 
Importantly, the result can be systematically improved by increasing $\mathfrak n$.

Next, let us benchmark the training of the ANN.
For this purpose we ideally would like to check the overlap $F=| \braket{ \psi_{ \mathcal{W}^{ (n) } }(t) }{ \psi_{QC}(t) } |$ of the final ANN-state to the one obtained from an exact application of the quantum circuit, which is impossible for large system sizes.
Nevertheless, we can offer a lower bound $F^\ast = \prod_k F_k^\ast<F$ where $F_k^\ast=| \bra{ \psi_{ \mathcal{W}^{ (k+1) } }(t) } U^{(k)}(t) \ket{ \psi_{ \mathcal{W}^{ (k) } }(t) } |$ denotes the partial overlaps measured at the end of each iteration $k$, which are a by-product of the training procedure.
We plot $F^\ast$ in Fig. \ref{fig:benchmarks} as a function of time for 1D- and 2D-lattices. 
It shows a high, macroscopic overlap even for large system sizes and a systematic improvement on adding more units and hidden layers to the ANN.
From this finding we conclude that the quantum circuit can be applied essentially numerically exactly on the ANN.
For comparison, we also plot the result of a perturbative treatment
$\ket{\psi_{QC}(t)} \approx e^{-i H_0^{(n)} t} \sum_{\vec s} \prod_k \exp \qty( \bra{\vec s} S_k -S_k(t) \ket{\psi_0} ) \ket{\vec s}$,
i.e. a cumulant expansion of the quantum circuit, that neglects (higher-order) commutators between different $S_k$. 
It shows a rapid decay and thus confirms the circuit's non-perturbative nature.
In the appendix we show further benchmarks of the whole framework for a large integrable system.

\textit{Numerics.} As an application of our framework we now explore non-equilibrium dynamics involving global quenches that has been difficult to access so far in the large system size and long-time limit. 
\begin{figure}						  	
	\includegraphics[width=1\columnwidth]{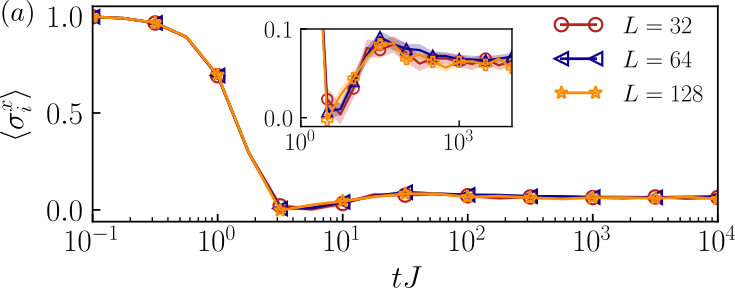}		
	\includegraphics[width=1\columnwidth]{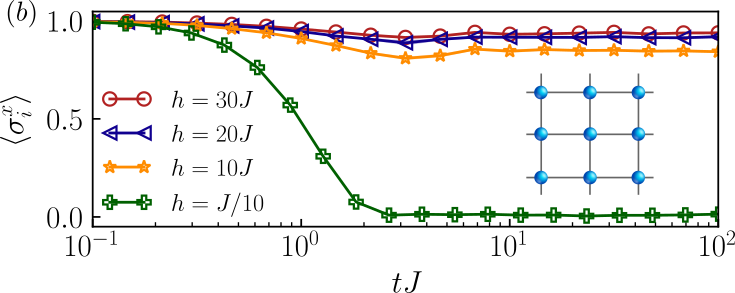}		
	\caption{Quench dynamics for the transverse magnetization and large system sizes averaged over lattice sites and 25 disorder realizations. (a) Quench into the MBL-SG phase at $h = J/4, J^{ (x) } = J/8$ for various system sizes. (b) $12\times 12$ lattice at very small and very large external field strengths.}
	\label{fig:magnetization}
\end{figure}
It is known from previous RG-studies that a symmetry-broken state will keep a non-zero Edwards-Anderson order parameter in the long-time limit starting from symmetry-broken states if the system is in the MBL-spin glass (MBL-SG) phase \cite{vosk2014dynamical}.
Here, we aim to address the \emph{build-up} of spatiotemporal order starting from a $\mathbb{Z}_2$-symmetric state upon quenching into the MBL-SG phase.
We detect the spatiotemporal dynamics of the MBL-SG order via~\cite{javanmard2019accessing},
\begin{equation}\label{eq:def:chi}
\chi_{ij}(t) = \sum_{\nu=1}^4 p_{ij}^{(\nu)}(t) \ev{\sigma_i^z \sigma_j^z}{\varrho_{ij}^{(\nu)}(t)}^2,
\end{equation}
where $\varrho_{ij}$ denotes the reduced density matrix of two lattice sites $i, j$, while $\nu$ enumerates its four eigenvectors $|\varrho_{ij}\rangle^{(\nu)}$ and eigenvalues (probabilities) $p_{ij}^{(\nu)}$. 
Fixing a distance $|i-j|$ we average $\chi_{ij}(t)$ across all associated pairs and disorder realizations.
This quantity can be interpreted as a local version of the Edwards-Anderson order parameter, which is otherwise mostly used to detect MBL-SG order in a static context, but which doesn't exhibit a natural extension to the dynamical regime considered here. 



Figure \ref{fig:quantum_circuit}b shows $\chi_d(t)$ both for an interacting MBL ($J^{(x)}=h/5$) and a non-interacting Anderson localized ($J^{(x)}=0$) case for a 1D chain of $64$ spins. 
At short times $tJ \tilde\le J/J^{(x)} = 25$, an almost identical light-cone for the buildup of MBL-SG correlations is visible, which appears consistent with a logarithmic growth.
On longer time scales we observe a fundamental difference between the Anderson and MBL cases.
For the non-interacting Anderson-localized limit the growth of MBL-SG order stops, while for $J^{(x)} > 0$ a second light-cone arises at a timescale that we estimate as $\sim  1 / J^{(x)}$.
Interestingly, we find that all light-cones do not become more open as we quench deeper into the MBL-SG phase but rather the more close we quench to the critical point.
This behavior is reminiscent of the $l$-bit picture, where LIOMs become more extended on approaching criticality. 
We will draw a connection to this picture below.
As expected, a quench within the MBL-PM phase does not show any SG-order.
Right at criticality, $J=h$, even without interaction, we find that the order becomes genuinely long-range as it decays algebraically with distance within the light-cone.
For the interacting case, inside the SG-phase, we observe an exponential decay with distance, but having an essential difference to the non-interacting case: the order at any fixed distance does not saturate, but increases strictly monotonically for all observed times within the light-cones.
This is a drastic non-perturbative effect of the interacting model.
It is particularly obvious for next-neighboring spins, see Fig.\ref{fig:quantum_circuit}b).
The important question whether this growing will eventually lead to a finite plateau for $|i-j|\to \infty$ requires access to even much later times, which we currently cannot access.

When initializing the system in a symmetry-broken state, as studied in previous works, the stability of MBL-SG order originates from the large overlap with the LIOMs.
The mechanism for the build-up of long-range order from symmetric states as targeted in this work is of fundamentally different origin, as the initial state is oriented orthogonal to the LIOMs.
%
Here, it is essential to generate long-distance quantum correlations between LIOMs.
This is not possible in the Anderson localized limit because the LIOMs are independent, as we also see from our results in Fig.~\ref{fig:quantum_circuit}.
Only in the interacting MBL limit the MBL-SG order can develop.
Quantum correlations between two lattice sites $i$ and $j$ can emerge on a time scale $[J^{(x)}]^{-1} e^{ |i-j|/\xi}$ where $\xi$ denotes a typical localization length.
Consequently, at a given time $t$ MBL-SG order can be generated over distances $d \sim \xi \log[J^{(x)}t]$ explaining the appearance of the logarithmic light-cone in Fig.~\ref{fig:quantum_circuit}.

As a closing point, we now turn briefly to quantum many-body dynamics in two dimensions. 
Whether a nonergodic phase due to strong disorder exists there has remained an outstanding challenge \cite{bordia2017probing}. 
Its difficulty originates from the percolation of many-body resonances \cite{de2017many, alet2018many}.
We find that at least for sufficiently small or large external fields, the latter can be effectively captured using our framework up to an unprecedented long timescale.
Fig. \ref{fig:magnetization}b shows the temporal evolution of the local magnetization in a quadratic, rectangular lattice, using essentially the same quench protocol as above.
In contrast to the glassy dynamics of a chain, see Fig. \ref{fig:magnetization}a, the lattice exhibits a rapid decay of magnetization at $h \ll J$, consistent with thermalization.
On the other hand, for $h \gg J$ a stable non-thermal plateau is reached.
Our result therefore numerically confirms a presumed quasi-localization \cite{de2017many, alet2018many} in the disordered 2D transverse-field Ising model at infinite temperature.

\textit{Conclusion.}
We have demonstrated how many-body quantum dynamics can be simulated for generic spin-1/2 systems up to exponentially long times given that sufficiently strong disorder breaks ergodicity at least up to the targeted timescale.
Importantly, this includes an \emph{unbiased} treatment of many-body resonances, which allowed us to obtain quantitative results in general and to go beyond one-dimensional systems.
%
We could show that our proposed framework does not fundamentally rely on any specific details of the model and scales up to systems sizes far beyond of what is possible with exact diagonalization.
This opens up for broad investigations e.g. of non-thermal behavior and quantum aging dynamics in higher dimensions \cite{wahl2019signatures, choi2016exploring}, long-range interacting systems \cite{zeiher2017coherent, piccitto2019dynamical, hauke2013spread} or localization in lattice gauge theories \cite{karpov2020disorder}.
Since this work has shown that deep ANNs are able to apply the proposed quantum circuit numerically exact, the ansatz could also be well suited for random unitary circuit models e.g. to study operator spreading \cite{nahum2018operator} \cite{von2018operator} \cite{khemani2018operator} or measurement induced localization transitions \cite{bao2020theory} \cite{jian2020measurement}.

%





{\it Acknowledgments.}
We are grateful to M. Schmitt, M. Schir{\' o}, G. De Tomasi and M. Schulz for helpful discussions. This project has received funding from the European Research Council (ERC) under the European Union’s Horizon 2020 research and innovation programme (grant agreement No. 853443), and M. H. further acknowledges support by the Deutsche Forschungsgemeinschaft via the Gottfried Wilhelm Leibniz Prize program. Moreover,the authors gratefully acknowledge the Gauss Centre for Supercomputing e.V. (www.gauss-centre.eu) for funding this  project  by  providing  computing  time  through  the John  von  Neumann  Institute  for  Computing  (NIC)  on the  GCS  Supercomputer  JUWELS  at  J{\"u}lich  Supercomputing Centre (JSC).

\bibliographystyle{apsrev4-1}
\bibliography{references} 

\appendix

\section{Comparison to exact solution at large scale}
\begin{figure}[h]
	\includegraphics[width=1\columnwidth]{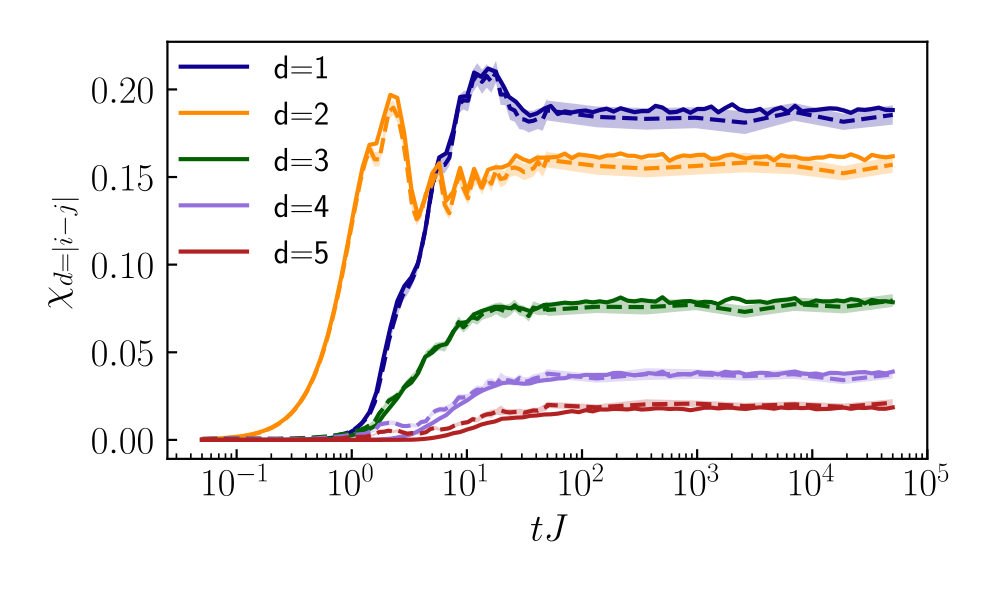}
	\caption{Comparison of spin-glass order computed with RG and ANN (dashed) to the exact solution (solid) for the non-interacting Ising chain for $L = 64, J = 5, h = 1$, at various distances $d$.}
	\label{fig:esg_benchmark}
\end{figure}

For the one-dimensional and non-interacting case of $J^{(x)} = 0$, $\chi_{|i - j|}(t)$ is exactly solvable by means of conventional free-fermion techniques \cite{van1980note, calabrese2011quantum} after performing a Jordan-Wigner transformation \cite{coleman2015introduction}.
We take this exact solution to compare our numerical result obtained from the prescribed framework at large system sizes, see Fig. \ref{fig:esg_benchmark}.
It shows an excellent agreement at the full range of timescales.

\section{Structure of the artificial neural network}
We use a complex-valued feed forward network taking a spin-configuration $\vec s$ as input layer and returning the activation of a single output unit as $H_{\rm ANN}(\mathcal{W}, \vec s) = \log[\psi_{\mathcal{W}}(\vec s)]$.
In between those, there are one or more hidden layers, each passing the previous, weighted and biased activations $W^{(\nu)} \vec v^{(\nu - 1)} + \vec b^{(\nu)}$ through a non-linear activation function $f(z)$ to the next layer, $v_j^{(\nu)} = f \qty(\sum_i W_{ij}^{(\nu)} v_i^{(\nu - 1)} + b_j^{(\nu)})$.
Building upon the original ansatz in terms of an RBM \cite{carleo2017solving}, we take the complex $\log\cosh(z)$ as a natural choice for the activation function in all hidden layers.
In the special case of a \emph{single} hidden layer, both formulations are in fact equivalent,
\begin{equation}
\begin{split}
	\ket{\psi_{\mathcal W}} = \sum_{ \{ \vec s, \vec h \} } e^{\vec a \cdot \vec s + \vec b \cdot \vec h + \vec s \cdot W \cdot \vec h } \ket{\vec s}, \ \vec h = (h_2, h_2, \ldots), h_i = \pm 1 \\ = \sum_{ \{ \vec s \} } e^{ \vec a \cdot \vec s + \sum_j \log\cosh[ \theta_j(\vec s) ] } \ket{\vec s},\ \theta_j(\vec s) = \sum_i W_{ij} s_i + b_j,
\end{split}
\end{equation}
which justifies our choice, although we note that a formal way of deriving an optimal $f(z)$ does not exist in machine learning.
In the above definition, $\mathcal{W} = (\vec a, \vec b, W)$ summarizes all network parameters.

Unfortunately, however, by using $f(z)=\log\cosh(z)$ we frequently observe the occurrence of numerical instabilities during training, caused by two poles located at $\pm i\pi / 2$.
These instabilities are triggered whenever $z$ comes close to those poles. 
This manifests itself in sudden jumps of $L [ \mathcal{\tilde W}^{ (k) } ]$, which can ultimately make convergence impossible.
To fix this problem, we use an approximation $\tilde f(z) \approx \log\cosh(z)$ that "smooths" the poles while preserving the asymptotic behavior:
\begin{align}
\log\cosh(z) = -\log(2) + z + \log \qty( 1 + e^{-2z} ) \\
\approx -\log(2) + z + \frac{P_2(z)}{Q_4(z)} =: \tilde g(z).
\end{align}
Here, a Pad{\'e}-$(2, 4)$ expansion is done in the second line.
The four poles of $\tilde g(z)$ are all located within $\Re(z) < 0$, so we choose
\begin{equation}
\tilde f(z) =
\begin{cases}
	\tilde g(z), & \Re(z) \ge 0 \\
	\tilde g(-z), & \Re(z) < 0
\end{cases}
\end{equation}
to avoid the poles and make $\tilde f(z)$ an even function like $\log\cosh(z)$.
If $\tilde f(z)$ is used, no more instabilities occur.

\section{Continuous renormalization}
As already mentioned in the main text, a SWT with a generator $S$ fulfilling $[H_0, S] = V$ allows to separate $\tilde H_0$ up to second order. 
We refer to \cite{pekker2014hilbert} for a formal way to obtain $S$.
Here we recall that $H_0$ represents a single coupling of arbitrary type, e.g. $\sigma_i^x$ or $\sigma_i^z \sigma_j^y \sigma_k^y \sigma_l^z$.
Although the SWT can be extended to any order, there is a subtlety which far more limits the overall accuracy than its order which is the following.
We refer to couplings $H_0'$ which commute with $H_0$ but still produce new, non-commuting couplings $V'$ under the SWT of any order:
\begin{equation}
[H_0', H_0] = 0, \ [e^S H_0' e^{S^\dagger}, H_0] = [\tilde H_0' + V', H_0] \ne 0 .
\end{equation}
These new couplings $V'$ need to be damped in subsequent SWTs before $\tilde H_0$ is removed, unless they can be considered as irrelevant for specific models \cite{vosk2014dynamical}.
Since we aim for a general framework and quantitative dynamics, our strategy is to keep \emph{all} emerging couplings while performing a continuous unitary transformation (CUT), whereupon $\tilde H_0$ is removed.
To formalize this procedure, we define a continuous scale $\lambda \in [0, \infty)$, where for $\lambda \to \infty$, $\tilde H_0 = H_0(\lambda \to \infty)$ commutes with all other couplings, very much like in the flow equation method \cite{thomson2018time}.
Thereby, like in first order SWT, we require the generator to satisfy 
\begin{equation}\label{eq:contSW}
[ H_0(\lambda), S(\lambda) ] = \sum_{ \frac{| V_i(\lambda) |}{| H_0(\lambda) |} > \epsilon } V_i(\lambda),
\end{equation}
but only w.r.t. those non-commuting couplings $V_i(\lambda)$, whose relative magnitude lies above a threshold $\epsilon \ll 1$.
The CUT itself is given by
\begin{equation}
\begin{split}
\frac{d}{d\lambda} H(\lambda) &= [S(\lambda), H(\lambda)], \\
H(\lambda) &= H_0(\lambda) + H_0'(\lambda) + V(\lambda),
\end{split}
\end{equation}
which, under the condition of Eq.~\eqref{eq:contSW}, converges to
\begin{equation}
H(\lambda \to \infty) = \tilde H_0 + \tilde H_0' + \sum_{ | \tilde V_i | / | \tilde H_0 | < \epsilon } \tilde V_i .
\end{equation}
By tuning the threshold $\epsilon$, the number of new couplings emerging during the CUT can be controlled without the technical need to restrict their type, i.e. the associated Pauli-string, by any means.
In the limit of $\epsilon \to 0$, the separation of $\tilde H_0$ becomes exact.

From the CUT we numerically construct a finite sequence of SW-generators $(S_1, S_2, \ldots)$.
The chained sequences of all RG-steps form the total sequence of $\{ S_k \}$ referred to in the main text.
Its length can be further optimized by merging commuting consecutive elements.

\section{Local rotations of RG-generators}
Our definition of time-dependent unitaries $U_k(t) = e^{\tilde S_k^\dagger(t)} e^{\tilde S_k}$ requires RG-generators $S_k$ to be successively rotated into the frame of all previous ones,
\begin{equation}\label{eq:S_k_tilde}
\tilde S_k = e^{S_1^\dagger} \cdots e^{S_{k-1}^\dagger} S_k e^{S_{k-1}} \cdots e^{S_1},
\end{equation}
as shown in the main text.
Like the Hamiltonian, we represent all $S_k$ as sums of Pauli-strings but with an \emph{imaginary} coefficient each, to ensure $S_k^\dagger = -S_k$.
In the following, we refer to each element of these sums as coupling.
While in general Eq.~\eqref{eq:S_k_tilde} generates exponentially many couplings, we empirically find that those being smaller than the threshold $\epsilon$ can be neglected after each rotation. 
Such a repetitive cropping does not alter expectation values of local observables at arbitrary times, up to a sub-leading correction, see below.
We attribute this observation to a relatively low overlap among respective $S_l$.
However, we strongly emphasize that this would not hold if the generators were replaced by $U_l(t)$, i.e. the quantum circuit itself. 
For intermediate to long times, all $U_l(t)$ have an extended spatial support implying a \emph{high} overlap among them, as illustrated in Fig. \ref{fig:quantum_circuit}a.
This quickly leads to an explosion of non-negligible couplings during successive rotations as we confirm in numerical experiments and by the rapid decay of a cumulant expansion approach as shown in Fig. \ref{fig:benchmarks}b.
Instead, these higher-order and long-distant couplings can be numerically exactly captured using a deep ANN, which is a central result of the present work.

As a sub-leading correction we impose the Frobenius norm of an original generator $S_l$ to the associated cropped generator $S'_l$,
\begin{equation}
\bar S'_l = \frac{|| S_l ||_F}{|| S'_l ||_F} S'_l,\quad S'_l = \sum_{| S_l^{(i)} | > \epsilon} S_l^{(i)},
\end{equation}
where $S_l^{(i)}$ denotes a coupling within $S_l$.


\end{document}